\documentclass{elsart}

\usepackage{epsfig}
\usepackage{amsmath}
\usepackage{amssymb}
\usepackage{latexsym}
\usepackage{exscale}

\renewcommand{\vec}[1]{\mbox{\boldmath{$#1$}}}

\begin{document}

\begin{frontmatter}

\title{Gas gain and signal length measurements with a 
triple-GEM at different pressures of
Ar-, Kr- and Xe-based gas mixtures\thanksref{EU}}
\thanks[EU]{Work supported by the European Community
            (contract no. ERBFMGECT980104).}

\author[Siegen]{A. Orthen\corauthref{Autor}},
\author[Siegen]{H. Wagner},
\author[Siegen]{H.J. Besch},
\author[Siegen]{S. Martoiu},
\author[Trieste]{R.H. Menk},
\author[Siegen]{A.H. Walenta},
\author[Siegen]{U. Werthenbach}

\corauth[Autor]{Corresponding author.
Tel.: +49 271-740-3563;
fax:  +49 271-740-3533;
e-mail: orthen@alwa02.physik.uni-siegen.de.}

\address[Siegen]
  {Universit\"at Siegen, Fachbereich Physik,
   Emmy-Noether-Campus,
   Walter-Flex-Stra\ss e 3, 57068 Siegen, Germany}
\address[Trieste]
  {ELETTRA, Sincrotrone Trieste, S.S. 14, km 163.5,
   Basovizza, 34012 Trieste, Italy}

\begin{abstract}
We investigate the gas gain behaviour
of a triple-GEM configuration
in gas mixtures of argon, krypton and xenon
with ten and thirty percent of carbon dioxide at pressures between $1$
and $3\,\mathrm{bar}$. 
Since the signal widths affect the dead time behaviour of the detector 
we present signal length measurements to evaluate
the use of the triple-GEM in time-resolved X-ray imaging.
\\
\vspace{4mm}
\end{abstract}


\end{frontmatter}

\section{Introduction}
In the recent years the application range of GEMs (gas electron
multipliers) \cite{Sauli}
in detector development has been widened up. 
Besides the operation in detectors
at high energy projects like HERA-B \cite{Zeuner}, COMPASS
\cite{Ketzer} or DIRAC \cite{Adeva} the GEM
is also used in gaseous photomultiplier development 
\cite{Buzulutskov1,Buzulutskov2} or X-ray imaging
detectors \cite{Li,Bachmann2}. 
Thereby, the GEM is usually combined with other micro pattern 
devices like MSGCs \cite{Benhammou,Baruth}, 
MGCs \cite{Fonte}, micromegas \cite{Kane} 
or MicroCAT \cite{Orthen2}. Also the cascade of multiple GEMs
has been tested extensively \cite{Bachmann}.

A triple-GEM cascade working as a detector for solar neutrino and
dark matter search has been tested at high
pressure of pure noble gases \cite{Bondar1,Bondar2}. 
High pressure is also desired in photon detection to increase quantum
efficiency. If pure noble gases are not explicitly needed it is
helpful to add a quench gas fraction of up to some $10\,\%$ to the noble
gases. This leads normally to smaller charge diffusion, shorter 
signal lengths and the possibility to obtain higher gas gains.
Examinations of the gas gaining behaviour of alternative
micro pattern devices like MSGCs \cite{Shekhtman} 
or the behaviour of a single GEM/MSGC-combination \cite{Bondar3}
in noble gas/quench gas environments
show that the maximum gain drops with pressure.

The aim of our investigations is to
test the performance of a triple-GEM detector 
at higher pressures of noble gas/quench gas mixtures to increase 
the quantum efficiency for photon detection in the energy range of
$8$--$24\,\mathrm{keV}$. We investigate the gas gain behaviour
at different pressures and quench gas fractions. 
We examine the signal lengths which mainly affect the dead
time. Besides space charge effects the dead time limits the rate
capability.

\section{Detector setup}

\begin{figure}
 \vspace{0mm}
 \begin{center}
  \includegraphics[clip,width=7.5cm]{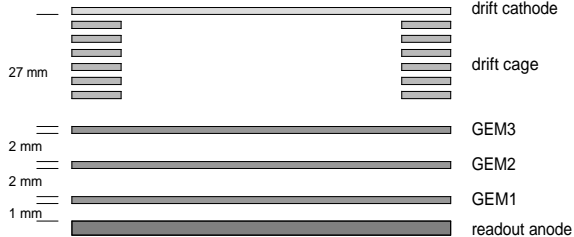}
 \end{center}
  \caption{Schematic cross section of the triple-GEM detector.}
  \label{fig_triplegem}
\end{figure}

The schematic setup of the triple-GEM detector is shown in
Fig. \ref{fig_triplegem}. 
The individual GEM foils with dimensions of
$6\times 6\,\mathrm{cm^2}$ are glued under slight tension to a
$1\,\mathrm{mm}$ thick ceramic frame.  The GEM type is characterised
by a hole pitch of $140\,\mathrm{\mu m}$ and an optical transparency
of $12\,\%$; the hole shape is double-conical with an inner diameter
of about $50\,\mathrm{\mu m}$ and an outer diameter of about
$90\,\mathrm{\mu m}$. The holes are hexagonally arranged;
the thickness of the GEM amounts to
$60\,\mathrm{\mu m}$ including two copper layers of $5\,\mathrm{\mu m}$
each and a Kapton thickness of $50\,\mathrm{\mu m}$.

We have set the distance between the GEMs to $2\,\mathrm{mm}$. The
lowermost GEM is mounted at a distance of $1\,\mathrm{mm}$ above the
readout structure. Following the usual nomenclature we denote the applied 
electric field between the GEMs by \emph{transfer field} and the
field between the lowermost GEM and the readout structure by
\emph{induction field}. Above the GEM configuration a drift cage is
mounted which provides a homogeneous \emph{drift field} in the conversion
gap of $27\,\mathrm{mm}$.

For the gas gain measurements (s. section \ref{sec_gasgain})
a copper anode with a size corresponding to the GEMs has been
used.
The signal examination (s. section \ref{sec_signals})
has been carried out with a
smaller copper anode ($\approx1\,\mathrm{cm^2}$) 
to reduce the capacitance and therefore the noise of the preamplifier.

\begin{figure}
 \vspace{0mm}
 \begin{center}
  \includegraphics[clip,width=7.5cm]{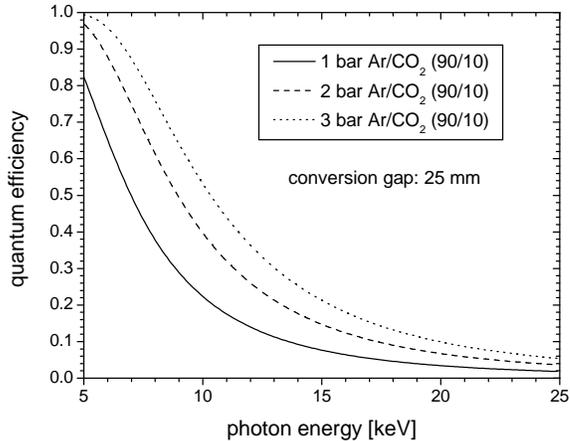}
 \end{center}
  \caption{Quantum efficiency as a function of photon energy 
 at pressures between $1$ and
 $3\,\mathrm{bar}$ of $\text{Ar/CO}_2$
 (90/10) with a conversion gap of $25\,\mathrm{mm}$.}
  \label{fig_cross_ar}
\end{figure}

\begin{figure}
 \vspace{0mm}
 \begin{center}
  \includegraphics[clip,width=7.5cm]{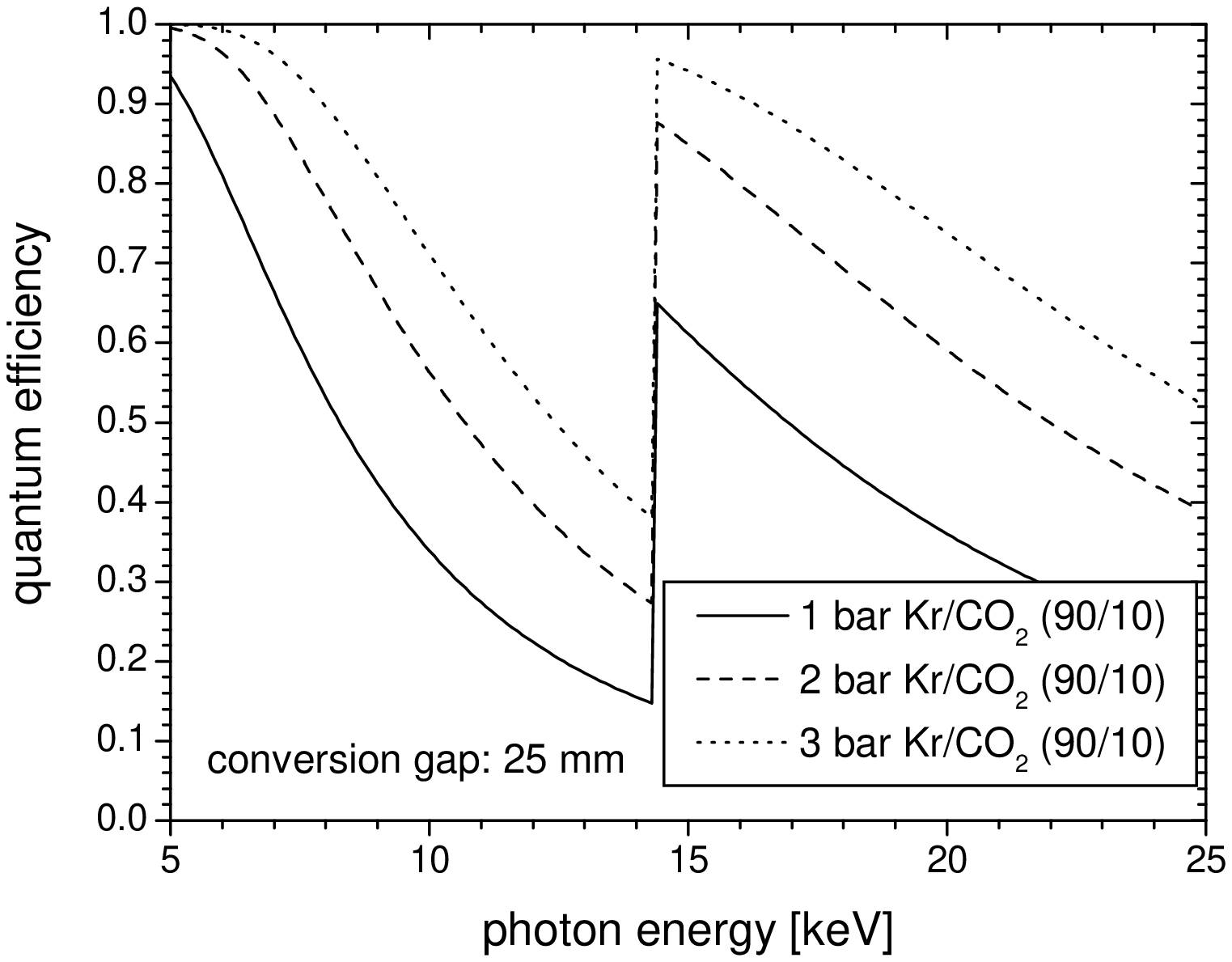}
 \end{center}
  \caption{Quantum efficiency  as a function of photon energy
   at pressures between $1$ and
 $3\,\mathrm{bar}$ of $\text{Kr/CO}_2$
 (90/10) with a conversion gap of $25\,\mathrm{mm}$.}
  \label{fig_cross_kr}
\end{figure}

\begin{figure}
 \vspace{0mm}
 \begin{center}
  \includegraphics[clip,width=7.5cm]{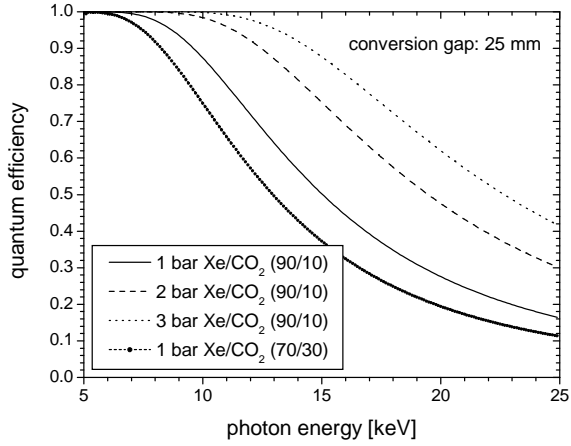}
 \end{center}
  \caption{Quantum efficiency  as a function of photon energy
   at pressures between $1$ and
 $3\,\mathrm{bar}$ of $\text{Xe/CO}_2$
 (90/10) and $1\,\mathrm{bar}$ of $\text{Xe/CO}_2$
 (70/30) with a conversion gap of $25\,\mathrm{mm}$.}
  \label{fig_cross_xe}
\end{figure}

The $1\,\mathrm{mm}$ thick carbon fibre entrance window of the
pressure vessel accepts pressures up to about $3\,\mathrm{bar}$. 
Figs. \ref{fig_cross_ar}--\ref{fig_cross_xe} show the quantum
efficiency of a detector filled with $\text{Ar/CO}_2$,
$\text{Kr/CO}_2$ and $\text{Xe/CO}_2$ (90/10)
with a typical conversion gap of $25\,\mathrm{mm}$.
To achieve
high quantum efficiency for photon detection in the energy range of
$5$--$14\,\mathrm{keV}$ xenon gas mixtures are preferably used. 
For photons in the energy range of about $14$--$24\,\mathrm{keV}$ 
quantum efficiency increases by the use of krypton
with a K-shell energy of $14.3\,\mathrm{keV}$. 
Higher quench gas fractions, showed as an example for $\text{Xe/CO}_2$
(70/30) in Fig. \ref{fig_cross_xe}, decrease
quantum efficiency. 

\section{Gas gain measurements\label{sec_gasgain}}

In the following section we present results of gas gain
measurements in the triple-GEM configuration. 
The effective gain $G_\text{eff}$ is experimentally determined 
by the quotient of the measured current at the
anode $I_\text{anode}$ and the ``incoming'' current $I_0$:
\begin{align}
G_\text{eff}=\frac{I_\text{anode}}{I_0}\,\,.
\end{align}
The current 
\begin{align}
 I_0=R\,e\,\frac{E_\gamma}{W}
 \label{eq_i0}
\end{align}
is determined by the conversion rate $R$ of the X-ray photons 
in counting mode of the detector,
by the photon energy $E_\gamma$ and by the average energy required to
produce one electron-ion pair in the gas $W$; $e$ denotes the elementary
charge.

\begin{figure}
 \vspace{0mm}
 \begin{center}
  \includegraphics[clip,width=7.5cm]{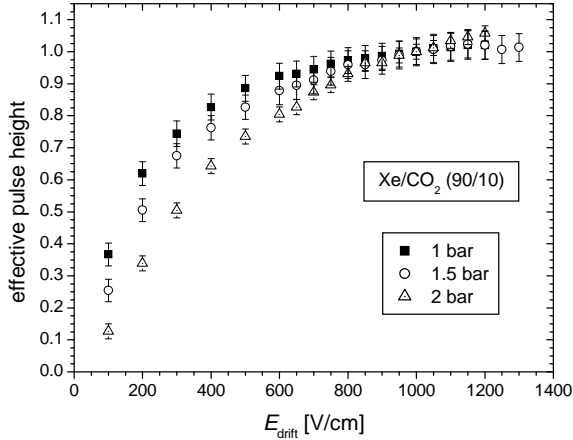}
 \end{center}
  \caption{Effective pulse height as a function of the applied drift
 field in $\text{Xe/CO}_2$ (90/10) for a transfer field of
 $2500\,\mathrm{V/cm}$ and an induction field of $3000\,\mathrm{V/cm}$.}
  \label{fig_dxe90}
\end{figure}

In fact, the incoming current $I_0$ is overestimated -- at least at 
small drift fields and high gas pressures --
because of charge losses due to recombination and attachment in
the conversion region.
This charge loss effect, which is mainly a three-body process and therefore
increases with the square of the gas pressure, is shown as an example 
in Fig. \ref{fig_dxe90}
in a $\text{Xe/CO}_2$
(90/10) gas mixture at pressures between $1$ and $2\,\mathrm{bar}$.
At drift fields larger than about $1000\,\mathrm{V/cm}$ the effective
pulse height reaches a plateau. In this region attachment and
recombination is negligible and the incoming current $I_0$ is described
correctly by Eq. (\ref{eq_i0}).

Since the maximum voltage applied to the drift
cathode was restricted by the detector geometry
to around $-(6000\pm500)\,\mathrm{V}$ we have chosen a drift field of
$400\,\mathrm{V/cm}$ to prevent the detector from sparking at the
drift cathode and to have enough margin to
increase more crucial potentials like GEM operation voltages.
The transfer fields have been chosen to $2500\,\mathrm{V/cm}$, and we have
applied an induction field of $3000\,\mathrm{V/cm}$ to avoid full
discharge propagation \cite{Bachmann3}. Furthermore, 
we have set the GEM voltages such,
that the uppermost GEM voltage is $10\,\%$ higher than the voltage of
the middle GEM; the lowermost GEM is set to an operation voltage 
$10\,\%$ smaller than the ``reference'' voltage  of the middle
GEM $\Delta U_\text{GEM2}$. This setting assures minimum discharge
rates \cite{Bachmann3}.
The sum of all applied GEM voltages always amounts to 
$3\,\Delta U_\text{GEM2}$. 

For illumination with X-ray photons we have used a $^{55}\text{Fe}$ 
source,
emitting photons with an energy of $5.9\,\mathrm{keV}$ and with a rate
of $41\,\mathrm{kHz}$ collimated to about $28\,\mathrm{mm^2}$.
During all measurements the gas gain was found to decrease 
as a function of illumination time and photon rate
most likely due to charging of the
insulating Kapton in the GEM. The charging
is dependent on illumination time and position, 
photon flux and dryness of the Kapton surface. We can not
explain why in our setup the gain decreased as a function of time,
whereas gain increased in measurements of other groups
(i.e. \cite{Bachmann},\cite{Bouclier}). The problem of
charging of the Kapton and
hence of the time and rate dependent gas gaining behaviour may be
solved by using carbon coated gas electron multipliers \cite{Beirle}.
In our case, however, all data has been taken
after an equilibrium state had been reached.

The gas gain measurements were limited by sparks. We stopped the
measurements at spark rates $\gtrsim2\,\mathrm{sparks/min}$, which is rather
an arbitrary limit. Important, however, is the stable detector
operation at slightly lower voltages and hence lower gas gains.  


\subsection{Gas gain in $\text{Ar/CO}_2$ (70/30)}

\begin{figure}
 \vspace{0mm}
 \begin{center}
  \includegraphics[clip,width=7.5cm]{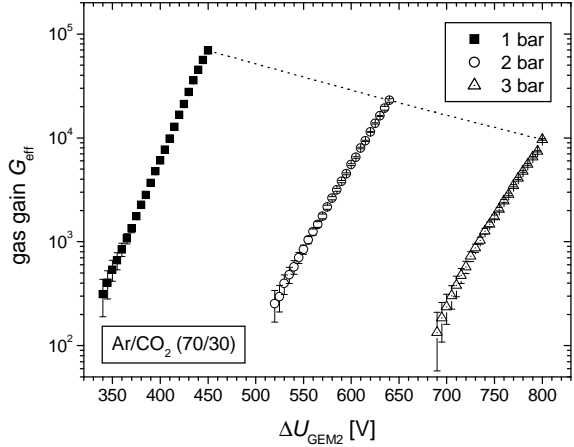}
 \end{center}
  \caption{Effective gas gain as a function of the applied voltage at the
  middle GEM $\Delta U_\text{GEM2}$ in $\text{Ar/CO}_2$ (70/30) at
  pressures of $1$, $2$ and $3\,\mathrm{bar}$. The voltages applied to
 the first and third GEM have been changed proportionally.}
  \label{fig_ggar70}
\end{figure}

Although quantum efficiency in argon mixtures for photon detection in
the higher energy range is small compared to krypton or xenon mixtures
we have carried out gas gain measurements in $\text{Ar/CO}_2$ (70/30)
to compare the results to those
achieved with Kr- and Xe-mixtures. The measured effective gain,
shown in Fig. \ref{fig_ggar70},
exceeds $6\cdot10^4$ at standard pressure. 
The maximum gas gain achievable decreases with pressure, but we have still
obtained a maximum gain of about $10^4$ at a pressure of
$3\,\mathrm{bar}$. 


\subsection{Gas gain in $\text{Kr/CO}_2$ mixtures}

\begin{figure}
 \vspace{0mm}
 \begin{center}
  \includegraphics[clip,width=7.5cm]{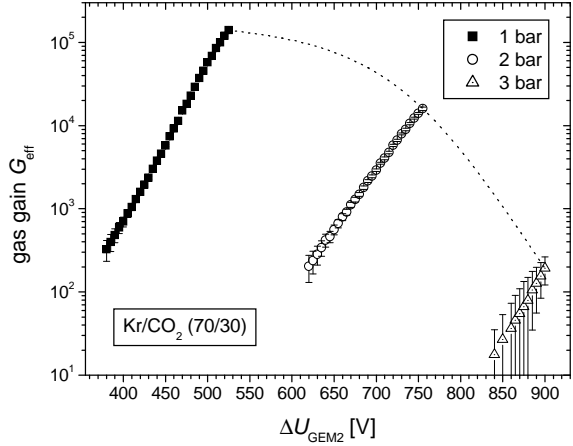}
 \end{center}
  \caption{Effective gas gain as a function of the applied voltage at the
  middle GEM $\Delta U_\text{GEM2}$ in $\text{Kr/CO}_2$ (70/30) at
  pressures of $1$, $2$ and $3\,\mathrm{bar}$.}
  \label{fig_ggkr70}
\end{figure}

\begin{figure}
 \vspace{0mm}
 \begin{center}
  \includegraphics[clip,width=7.5cm]{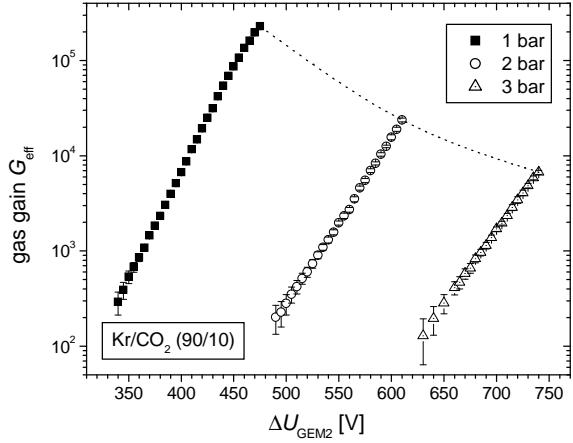}
 \end{center}
  \caption{Effective gas gain as a function of the applied voltage at the
  middle GEM $\Delta U_\text{GEM2}$ in $\text{Kr/CO}_2$ (90/10) at
  pressures of $1$, $2$ and $3\,\mathrm{bar}$.}
  \label{fig_ggkr90}
\end{figure}

We have investigated two Kr-mixtures with different $\text{CO}_2$
fractions of $30\,\%$ (see Fig. \ref{fig_ggkr70}) 
and $10\,\%$
(see Fig. \ref{fig_ggkr90}),
respectively. In both mixtures the effective gain exceeds $10^4$ up to
pressures of $2\,\mathrm{bar}$. Nevertheless, a quench gas fraction of
$30\,\%$ 
is not well suited for operation at pressures $>2\,\mathrm{bar}$ where
we obtained a strong decrease in gas gain to about $200$. In
contrast to that still large
gas gains in the order of $5\cdot 10^3$ are possible with
a carbon dioxide fraction of $10\,\%$. 


\subsection{Gas gain in $\text{Xe/CO}_2$ mixtures}

\begin{figure}
 \vspace{0mm}
 \begin{center}
  \includegraphics[clip,width=7.5cm]{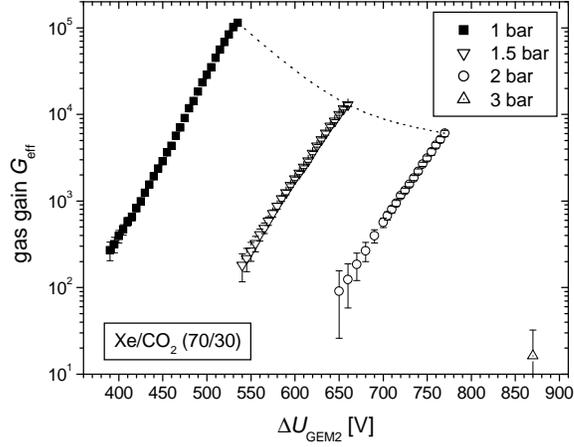}
 \end{center}
  \caption{Effective gas gain as a function of the applied voltage at the
  middle GEM $\Delta U_\text{GEM2}$ in $\text{Xe/CO}_2$ (70/30) at
  pressures of $1$, $1.5$, $2$ and $3\,\mathrm{bar}$.}
  \label{fig_ggxe70}
\end{figure}

\begin{figure}
 \vspace{0mm}
 \begin{center}
  \includegraphics[clip,width=7.5cm]{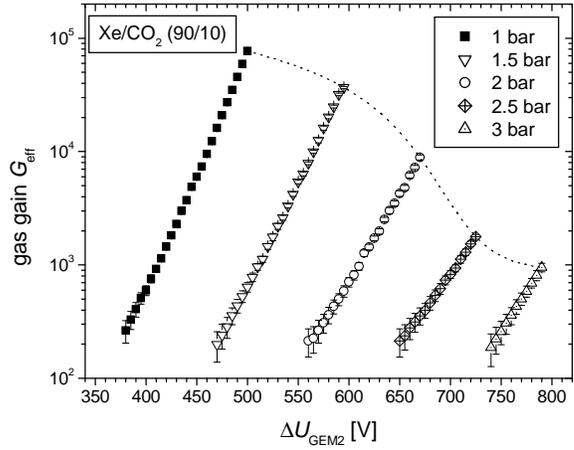}
 \end{center}
  \caption{Effective gas gain as a function of the applied voltage at the
  middle GEM $\Delta U_\text{GEM2}$ in $\text{Xe/CO}_2$ (90/10) at
  pressures of $1$, $1.5$, $2$, $2.5$ and $3\,\mathrm{bar}$.}
  \label{fig_ggxe90}
\end{figure}

The use of Xe-based gas mixtures offers the
largest quantum efficiency for photon detection 
in the energy range between $5$
and $14\,\mathrm{keV}$. Results of gas gain measurements for
$\text{Xe/CO}_2$ (70/30) and $\text{Xe/CO}_2$ (90/10) are shown in
Fig. \ref{fig_ggxe70} 
and Fig. \ref{fig_ggxe90}.
Like for Kr-gas mixtures an amount of $10\,\%$ carbon dioxide
enables operation with gas gains $>10^3$ at pressures above
$2\,\mathrm{bar}$. High quencher fractions are not well suited for
pressure operation with xenon.


\subsection{Homogeneity}

\begin{figure}
 \vspace{0mm}
 \begin{center}
  \includegraphics[clip,width=7.5cm]{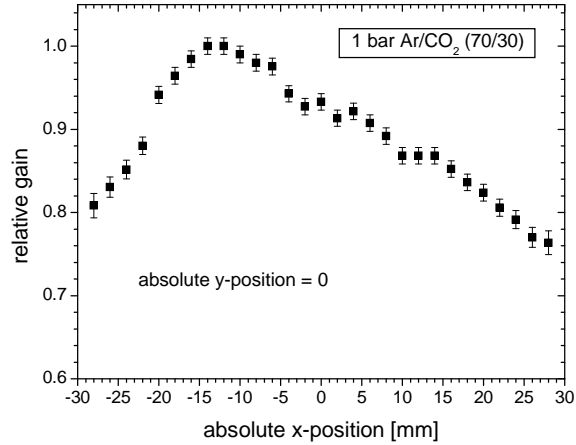}
 \end{center}
  \caption{Relative effective gain along the vertically centred
 $x$-axis of the detector in $\text{Ar/CO}_2$ (70/30) at standard
 pressure.}
  \label{fig_gainscan}
\end{figure}

Figure \ref{fig_gainscan} 
shows
the relative effective gain measured along the vertically centred 
$x$-axis. Gain variations of up to $25\,\%$ become obvious.
These local variations are mainly due to two aspects:
First, the GEMs, themselves,
produce intrinsic gain inhomogeneities because of geometry variations.
Furthermore, the distances between the individual 
GEMs and the distance between
undermost GEM and anode are not constant over all the detector area,
which leads to transfer and induction 
field inhomogeneities und finally to variations in the charge transfer
behaviour and in the effective gas gain of the GEMs \cite{Orthen2}.

\subsection{Discussion}

We have measured the gas gain behaviour at
different pressures of argon, krypton and xenon with additions of
$10\,\%$ and $30\,\%$ of carbon dioxide. 
A strong influence of the quench
gas fraction on the gas gain behaviour becomes obvious.
It is definitely a non-trivial task to
optimise the fraction of quencher at a given pressure with respect to a
maximum achievable gain due
to a complicated interplay of these parameters.

Generally, the maximum gas gain of the triple-GEM
achievable in slightly quenched noble gas 
mixtures is higher than in pure noble gases (compare to
Refs. \cite{Bondar1,Bondar2}).
At standard pressure
the krypton and xenon gas mixtures with $10\,\%$ $\text{CO}_2$ behave
relatively similarly to the mixtures with $30\,\%$
$\text{CO}_2$. Increasing gas pressure leads to a larger deviation in
maximum achievable gas gain.
Higher pressure demands obviously for smaller quench gas fractions. 

The strong influence of the quencher fraction indicates,
that a stable avalanche process is not only based on
electron-induced ionization of gas atoms. The UV-photons which are
produced in the avalanche development may contribute noticeably to the
multiplication procedure. With increasing pressure and constant (or
even larger) quench
gas fraction the mean free path of the UV-photons is decreasing.
For instance, the deterioration in energy 
resolution in a MWPC is stopped when a very small
quench gas fraction of only $2\,\%$ at very high pressures (up to
$30\,\mathrm{bar}$) is added to argon \cite{Feige}. However, still smaller
fractions of quenching gas lead to an unsafe operation of the
detector. We suppose, that the triple-GEM detector shows a comparable
behaviour. It should be considered to use very small quench gas fractions
of a few percent for really high pressure operation.

A good choice for photon detection with the triple-GEM is
the use of $\text{Kr/CO}_2$ or
$\text{Xe/CO}_2$ (90/10), depending on the energy range,
at pressures between $1$ and
$2\,\mathrm{bar}$. Gas gains of $5\cdot10^3$ can be easily reached
without sparking. 

\section{Signal length investigations \label{sec_signals}}

We have studied the dependence of the anode signal shape on GEM voltage,
induction field, transfer field and drift field for all gas mixtures
and several pressures since the dead time behaviour 
of the detector is mainly determined by the signals lengths.
The copper readout anode for these measurements
has a size of $1\times 1\,\mathrm{cm^2}$.
The integrating preamplifier and the shaper including a pole-zero filter
to cancel the pole of the integrating first stage are in-house
developments.
The gaussian-like $\delta$-response of this amplification system 
has a width of $15\,\mathrm{ns}$ (fwhm) due to bandwidth limitations.

\begin{figure}
 \vspace{0mm}
 \begin{center}
  \includegraphics[clip,width=7.5cm]{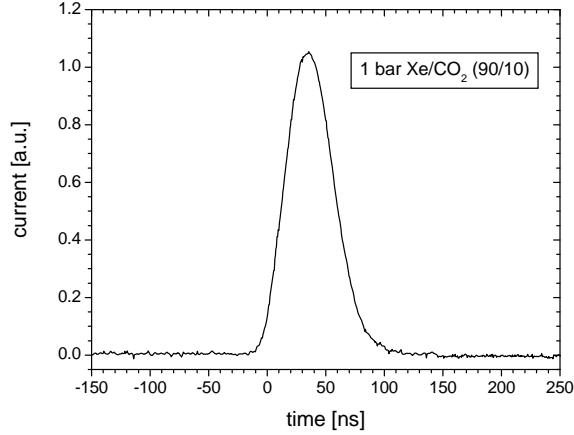}
 \end{center}
  \caption{Measured anode signal shape at a large drift field of
 $1000\,\mathrm{V/cm}$ in $1\,\mathrm{bar}$ 
  $\text{Xe/CO}_2$ (90/10).}
  \label{fig_ssxe90ed1000}
\end{figure}

\begin{figure}
 \vspace{0mm}
 \begin{center}
  \includegraphics[clip,width=7.5cm]{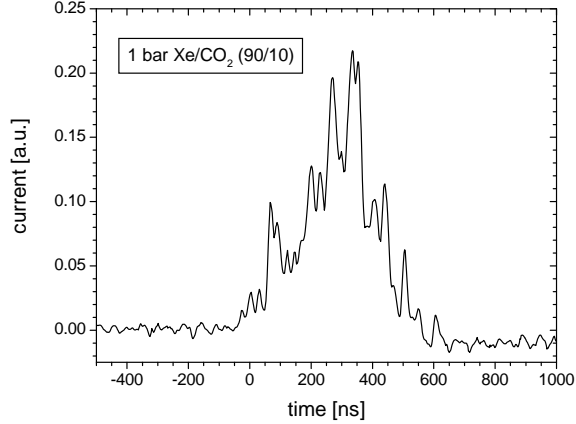}
 \end{center}
  \caption{Measured anode signal shape at a small drift field of
 $200\,\mathrm{V/cm}$ in $1\,\mathrm{bar}$ 
  $\text{Xe/CO}_2$ (90/10).}
  \label{fig_ssxe90ed200}
\end{figure}

Basically, the anode signal shape is determined by two aspects: 
\begin{enumerate}
\item{A spatially $\delta$-like distributed
electron cluster moving from the bottom side of the undermost
GEM towards the readout structure induces a box-shaped current at both 
involved electrodes 
if the electric
field and therefore the electron drift velocity 
in the region of interest is assumed to be constant \cite{Radeka}.
The copper-clad bottom side of the GEM shields all currents induced by
charge movements appearing above this electrode; especially the slow
ions do not contribute to the signal shape, which is, however, the
case for micromegas or MicroCAT detectors
\cite{Cussonneau,Sarvestani}. The
crucial parameters for the width $t_\text{w}$ of the box-shaped 
current are the electron drift velocity $v_\text{e}$
in the electric field of the induction region
and the distance $d$ between the bottom side of the GEM and
the anode which amounts to $1\,\mathrm{mm}$ in our case:
\begin{align}
t_\text{w}=\frac{d}{v_\text{e}}\,.
\label{eq_box}
\end{align}
}
\item{The longitudinal diffusion $\sigma_\text{l}$
of the primary produced electrons (photoelectron plus
subsequently produced about 300 electrons for $8\,\mathrm{keV}$
photons in argon) in the conversion region
leads to a temporal 
gaussian smearing 
of the above mentioned
box function, mathematically expressed by a convolution of box
and gauss function. 
The width of the gaussian
\begin{align}
\sigma_\text{t}=\left(\sigma_\text{l0}\,\overline{\sqrt{l}}\right)/
v_\text{e}
\label{eq_width}
\end{align}
is determined by the longitudinal diffusion coefficient 
$\sigma_\text{l0}$ and the electron drift velocity
$v_\text{e}$ in the
applied drift field and gas mixture used and finally by the mean
drift path $\left(\overline{\sqrt{l}}\right)^2$
of the primary produced
electrons, which is dependent on photon energy,
conversion gap and photoelectric absorption coefficient of the gas.
Additionally, the longitudinal electron
diffusion in the two transfer regions between the GEMs and in the
induction region between undermost GEM and anode contribute to the
temporal smearing $\sigma_\text{t}$. In our simulation
this contribution has also been taken into account.}
\end{enumerate}
For most gas mixtures and pressures in our detector setup the temporal
diffusion spreading (compare to Eq. (\ref{eq_width}))
is the main contribution which leads to a very much gaussian-like
signal shape.
Fig. \ref{fig_ssxe90ed1000}
shows a typical signal obtained at a high drift field of
$1000\,\mathrm{V/cm}$ at standard pressure. At smaller drift fields
the signal shape looks differently (s. Fig. \ref{fig_ssxe90ed200}).
The signal shape is not continuous anymore;
single electron clusters reach the anode one by one, but still the
gaussian character is well visible.
We have measured
the full width half maximum (fwhm) of the anode signals
and the $\delta$-response of the preamplifier
with a digital oscilloscope, afterwards
averaging the widths of several signals. 

Due to the gaussian signal shape and the gaussian $\delta$-response of
the preampli\-fier/shaper-system the real signal length can be
calculated to be the width of the deconvolution of both gaussians:
\begin{align}
\Delta_\text{sig-real}^2=\Delta_\text{sig-meas}^2-
\Delta_{\delta\text{-response}}^2\,,
\label{eq_deconv}
\end{align}
where $\Delta$ is the full width half maximum of the signal.
In case the measured signal shape is not gaussian (which means that the
main contribution of the signal is due to the
electron induced box-shaped current, which is true for very small
induction fields (s. section \ref{sec_induction})) 
Eq. (\ref{eq_deconv}) is no longer valid.
But in this case the measured signal width is very much larger
than the width of the $\delta$-response ($\Delta_\text{sig-meas} \gg
\Delta_{\delta\text{-response}}$) and therefore 
the real signal width can be
determined to be $\Delta_\text{sig-real}\approx\Delta_\text{sig-meas}$.

Despite most gas properties are $E/p$-dependent 
all measurements have been performed with a constant transfer field of
$2500\,\mathrm{V/cm}$ and a constant 
induction field of $3000\,\mathrm{V/cm}$ even at high pressure
if not noted otherwise, since these electric fields 
ensure maximum electron transparency of the GEM. 
The $^{55}\text{Fe}$ source has been collimated to about
$12\,\mathrm{mm^2}$; photons were emitted with a rate of a few hundred
$\mathrm{Hz}$. We have 
carried out all signal width simulations only for $1\,\mathrm{bar}$ 
$\text{Ar/CO}_2$ (70/30), where experimental data on drift velocities
\cite{Ma} and simulated data on longitudinal diffusion
\cite{Magboltz1,Magboltz2} have been used.

\subsection{Influence of GEM operation voltage}

We have detected no influence of the GEM voltage on the signal length.
The range of the investigated GEM voltages was limited by the dynamic
range of the preamplifier. Anyway, the setting of the GEM voltage is
determined by the desired gas gain and is therefore no freely
adjustable parameter for optimisation of the signal lengths.

\subsection{Influence of the induction field \label{sec_induction}}

\begin{figure}
 \vspace{0mm}
 \begin{center}
  \includegraphics[clip,width=7.5cm]{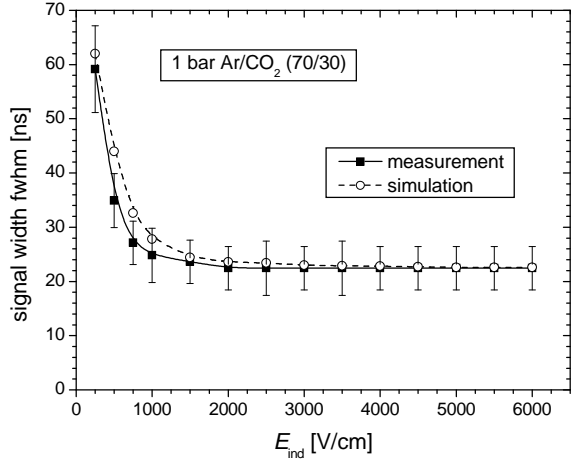}
 \end{center}
  \caption{Comparison between measured and simulated 
  signal widths (fwhm) as a function of the induction field
  in $1\,\mathrm{bar}$ 
  $\text{Ar/CO}_2$ (70/30) with a constant drift field of
  $E_\text{drift}=900\,\mathrm{V/cm}$.}
  \label{fig_slar70eind}
\end{figure}

The contribution of the width of the box-shaped current 
to the total width of the
signal increases for very small induction fields due to a
decreasing electron drift velocity (compare to Eq. (\ref{eq_box})). 
Three-dimensional
field calculations with
Maxwell \cite{Maxwell} using a geometry of $1\,\mathrm{mm}$ transfer
gap, the $60\,\mathrm{\mu m}$ thick GEM foil 
and $1\,\mathrm{mm}$ induction gap (the geometry of the
GEM and more simulation details are described in Ref. \cite{Orthen2})
show that the electric field in the induction
region along the symmetry axis of a GEM hole is larger
than the quotient $E=U_\text{GEM-bottom}/d$ of the applied 
voltage on the bottom side of the GEM $U_\text{GEM-bottom}$
and the distance $d$ between GEM bottom and anode.
This effect is some relict due to the very high
electric field in the holes. To determine the 
width of the electron induced box-shaped current we have used
a parametrisation of the electron drift velocity, calculated by the
Magboltz programm \cite{Magboltz1,Magboltz2}, up to electric fields of
$20\,\mathrm{kV/cm}$.
An average drift velocity in the induction region has been
calculated and the box function is convoluted with a gaussian,
describing the diffusion spreading. The
results of this simulation and the comparison to experiment is shown in
Fig. \ref{fig_slar70eind}.
A good agreement especially for induction fields $\ge
1\,\mathrm{kV/cm}$ becomes obvious. 
At induction fields larger than
about $2000\,\mathrm{V/cm}$ the influence of the box-shaped current
becomes
negligible compared to the contribution of the temporal electron 
diffusion in the conversion region.

\subsection{Influence of the transfer field}

\begin{figure}
 \vspace{0mm}
 \begin{center}
  \includegraphics[clip,width=7.5cm]{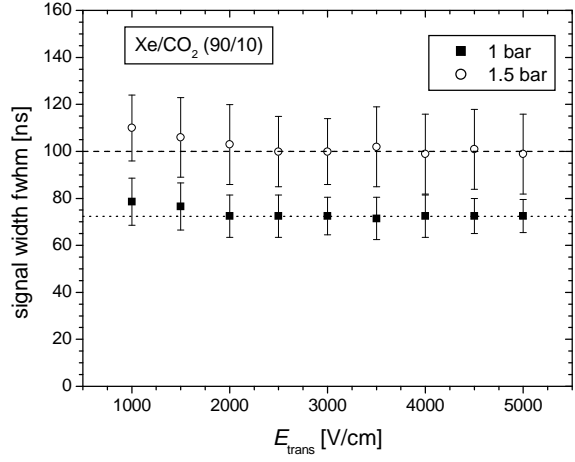}
 \end{center}
  \caption{Signal widths (fwhm) as a function of the transfer field
  between the uppermost and the middle and between the middle and the
  undermost GEM in $1$ and $1.5\,\mathrm{bar}$ 
  $\text{Xe/CO}_2$ (90/10) with a constant drift field of
  $E_\text{drift}=1000\,\mathrm{V/cm}$.}
  \label{fig_slxe90etrans}
\end{figure}

For this study both transfer fields between GEM1 and GEM2 and between
GEM2 and GEM3 have been changed at the same time.
The measured results are shown in Fig. \ref{fig_slxe90etrans}.
Transfer fields smaller than $2500\,\mathrm{V/cm}$ result in a slight
increase of the signal width which can be explained by 
smaller electron drift velocities in the transfer regions
and hence a larger diffusion spreading (compare to Eq. (\ref{eq_width})). 
At larger transfer fields the signal width reaches a plateau.
Higher pressure reflects smaller electron drift velocities and
therefore larger signal widths.

\subsection{Influence of the drift field}

\begin{figure}
 \vspace{0mm}
 \begin{center}
  \includegraphics[clip,width=7.5cm]{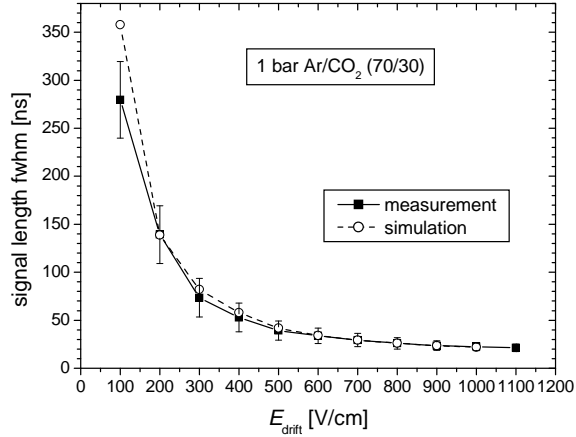}
 \end{center}
  \caption{Comparison between measured and simulated signal widths
  (fwhm) as a function of the drift field in $1\,\mathrm{bar}$ 
  $\text{Ar/CO}_2$ (70/30).}
  \label{fig_slar70sim}
\end{figure}

This section contains an intensive investigation of the influence of
the applied drift field on the signal width.
The comparison between measurement and simulation (see
Fig. \ref{fig_slar70sim})
shows a good agreement for drift fields $\gtrsim200\,\mathrm{V/cm}$. 
The signal width measurements at small
drift fields are very difficult due to the clustered signal shape
(see Fig. \ref{fig_ssxe90ed200}). 
Therefore, the error is relatively
large when determining these signal lengths. 
Furthermore, a large part of the primary produced electrons is
lost due to recombination and attachment at low drift fields (compare
to Fig. \ref{fig_dxe90}) and thus decreasing the signal-to-noise ratio. 

\begin{figure}
 \vspace{0mm}
 \begin{center}
  \includegraphics[clip,width=7.5cm]{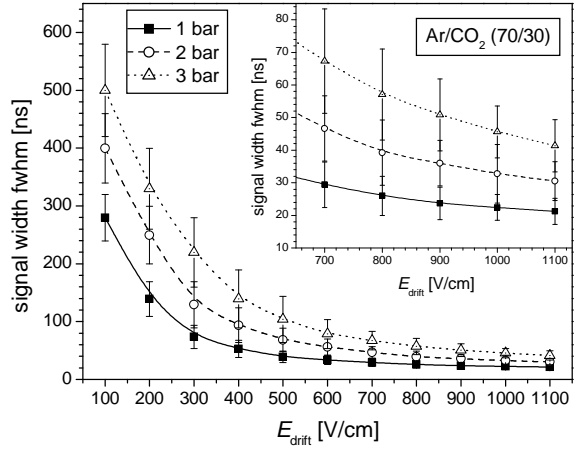}
 \end{center}
  \caption{Measured signal widths
  (fwhm) as a function of the drift field in 1,2 and $3\,\mathrm{bar}$ 
  $\text{Ar/CO}_2$ (70/30).}
  \label{fig_slar70}
\end{figure}

\begin{figure}
 \vspace{0mm}
 \begin{center}
  \includegraphics[clip,width=7.5cm]{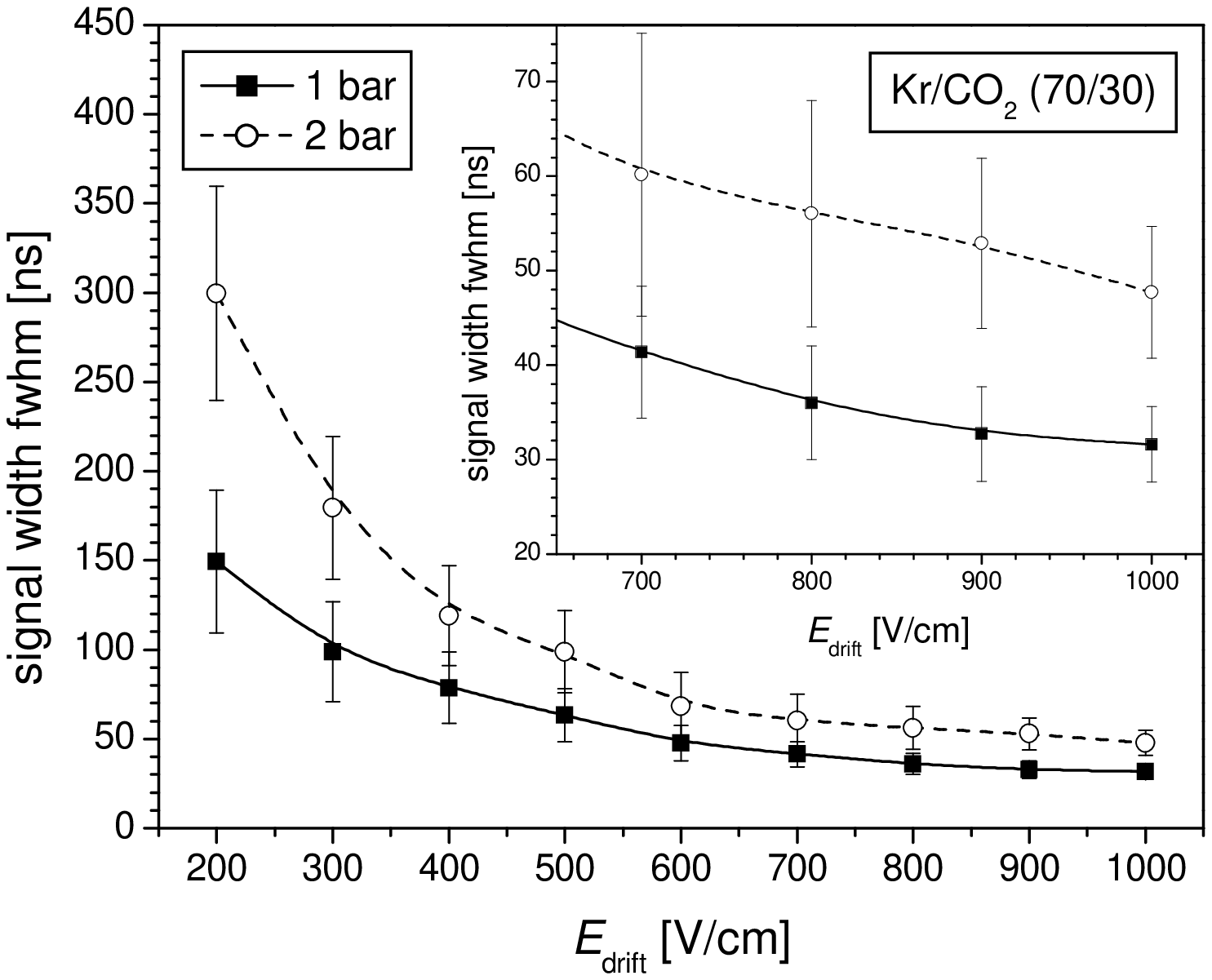}
 \end{center}
  \caption{Measured signal widths
  (fwhm) as a function of the drift field in 1 and $2\,\mathrm{bar}$ 
  $\text{Kr/CO}_2$ (70/30).}
  \label{fig_slkr70}
\end{figure}

\begin{figure}
 \vspace{0mm}
 \begin{center}
  \includegraphics[clip,width=7.5cm]{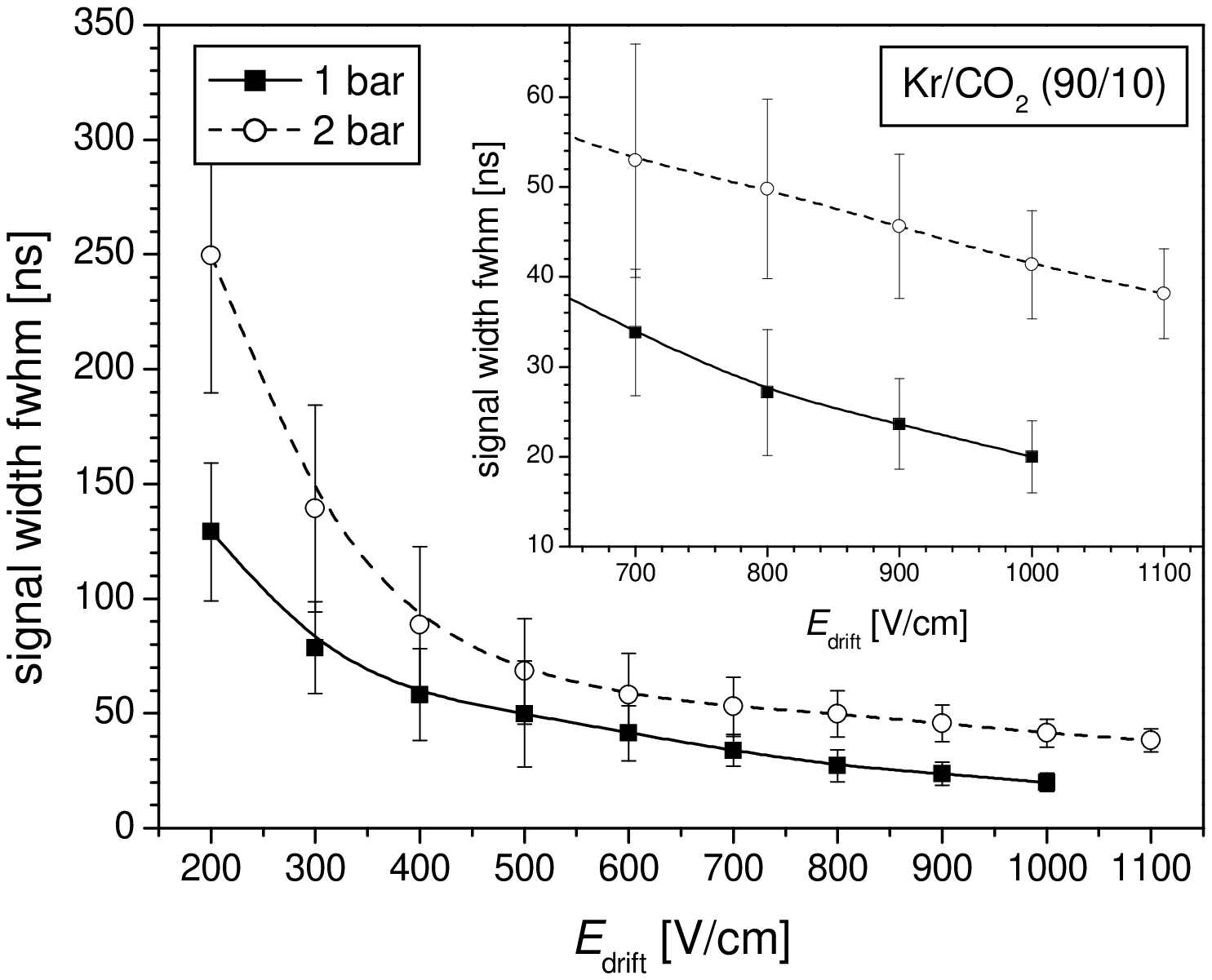}
 \end{center}
  \caption{Measured signal widths
  (fwhm) as a function of the drift field in 1 and $2\,\mathrm{bar}$ 
  $\text{Kr/CO}_2$ (90/10).}
  \label{fig_slkr90}
\end{figure}

\begin{figure}
 \vspace{0mm}
 \begin{center}
  \includegraphics[clip,width=7.5cm]{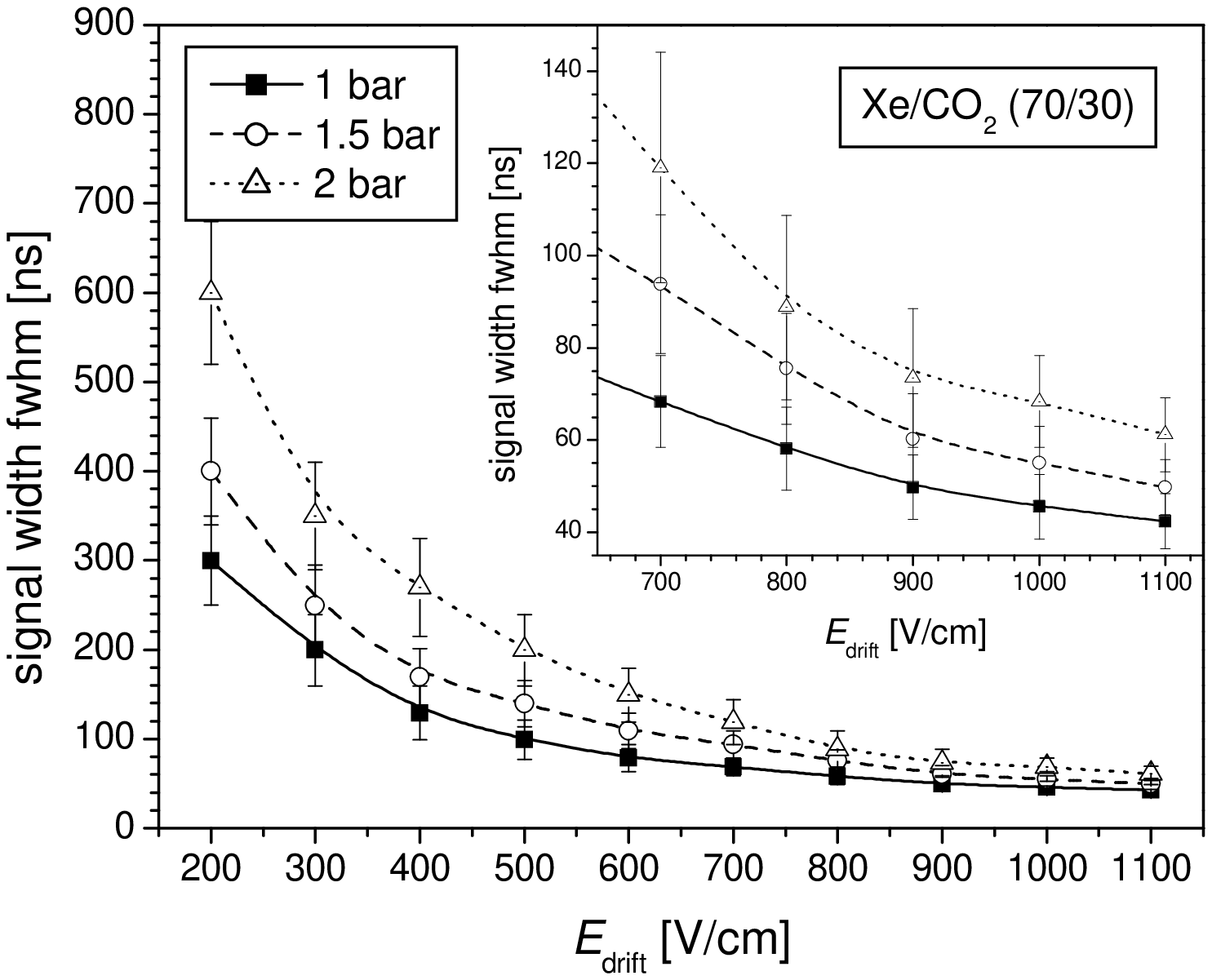}
 \end{center}
  \caption{Measured signal widths
  (fwhm) as a function of the drift field in 1, 1.5 and $2\,\mathrm{bar}$ 
  $\text{Xe/CO}_2$ (70/30).}
  \label{fig_slxe70}
\end{figure}

\begin{figure}
 \vspace{0mm}
 \begin{center}
  \includegraphics[clip,width=7.5cm]{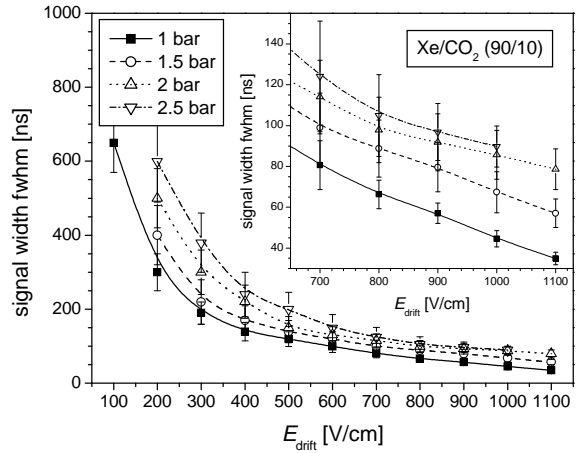}
 \end{center}
  \caption{Measured signal widths
  (fwhm) as a function of the drift field in 1, 1.5, 2  
  and $2.5\,\mathrm{bar}$ 
  $\text{Xe/CO}_2$ (90/10).}
  \label{fig_slxe90}
\end{figure}

The measured signal widths 
for different gas mixtures and gas pressures are shown 
in Figs. \ref{fig_slar70}--\ref{fig_slxe90}. 
Generally, high drift fields are needed to produce short signals. We
were limited to about $1000$--$1200\,\mathrm{V/cm}$ due to sparking at
the drift cathode.
The fastest signals with a signal length of $<25\,\mathrm{ns}$ (fwhm) 
are produced in the argon mixture at standard
pressure (see Fig. \ref{fig_slar70}). 
Larger pressure always increases the signal width due to
decreasing drift velocities in induction and conversion region 
and thus an increasing temporal smearing.
But even in $1.5\,\mathrm{bar}$ $\text{Xe/CO}_2$ (90/10)
the signal width does not exceed $60\,\mathrm{ns}$ (fwhm) at a drift field of
$1100\,\mathrm{V/cm}$. 

\subsection{Discussion}

In our detector setup the drift field has  
the most crucial influence on the signal length although for smaller
conversion regions this parameter becomes less important due to
smaller longitudinal electron diffusion. 
The drift field should be chosen as large as allowed by
the detector setup. However, the drift field should not be too high
($\lesssim2\,\mathrm{kV/cm}$) because the uppermost
GEM loses transparency for
too large drift fields \cite{Bachmann}. Transfer and induction fields 
$\gtrsim2\,\mathrm{kV/cm}$ minimise the widths of the signal. 
If, however, the contribution of the electron induced box-shaped current 
becomes
noticeable the distance between the undermost GEM and the anode can be
reduced, i.e. to $0.5\,\mathrm{mm}$. This should suppress this effect.
The GEM operation voltage, which is mainly determined by the desired 
gas gain, does not influence the signal widths visibly.

\section{Conclusion}
We have measured the gas gain in a triple-GEM geometry for pressures
up to $3\,\mathrm{bar}$ of $\text{Ar/CO}_2$ (70/30), 
$\text{Kr/CO}_2$ (70/30), $\text{Kr/CO}_2$ (90/10),
$\text{Xe/CO}_2$ (70/30) and $\text{Xe/CO}_2$ (90/10). At a pressure
$>2\,\mathrm{bar}$ the
addition of $10\,\%$ of carbon dioxide to the noble gases leads to
higher maximum gas gains compared to noble gases mixed with $30\,\%$
quench gas.
Too large amounts of quench gas seem to
disturb operation at high pressure and is therefore not
preferable. Generally, we have obtained that the quench gas fraction
should decrease with pressure.

Compared to the geometry of MicroCAT- or micromegas-detectors 
the triple-GEM-detector offers faster signals since only the
electrons and not the slow ions contribute to the signal shape at the
anode.
The measured signal lengths at high drift fields in all
gas mixtures and pressures
investigated are smaller than about $100\,\mathrm{ns}$
(fwhm). If transfer and induction fields exceed $2\,\mathrm{kV/cm}$
the signal widths get to minimum.

Provided that the signal width is mainly determined by the temporal
diffusion of the primary electrons in the conversion region
further optimisation with respect to an increased high rate behaviour
can be performed with exact knowledge of photon energy and desired
quantum efficiency by reducing the conversion gap (compare to
Figs. \ref{fig_cross_ar}--\ref{fig_cross_xe}).
For example, with a desired quantum efficiency of
$90\,\%$ for $8\,\mathrm{keV}$-photons in $1.5\,\mathrm{bar}$
$\text{Xe/CO}_2$ (90/10) it is sufficient to provice a conversion gap 
of $11\,\mathrm{mm}$ with mean drift path 
$\left(\overline{\sqrt{l_{11}}}\right)^2$
instead of $27\,\mathrm{mm}$ with mean drift path
$\left(\overline{\sqrt{l_{27}}}\right)^2$
used the for measurements described in this paper.
Hence, the longitudinal diffusion and consequently the signal width
will decrease by a
factor of $\overline{\sqrt{l_{27}}}\,/\,\overline{\sqrt{l_{11}}}
\approx 1.8$.

In combination with the resistive readout structure \cite{Besch}
it is proposed to
integrate about $90\,\%$ of the charge to achieve optimum spatial
resolution \cite{Hendrik1}. 
Compared to the signal length fwhm, presented here, the
charge integration time from $5\,\%$ to $95\,\%$ of the
original signal can be calculated to be
$\tau_{5-95\,\%}\approx1.4\,\tau_\text{fwhm}$. Due to charge 
diffusion on the readout structure the charge integration time for
signals with a width of $\tau_\text{5-95\,\%}=92\,\mathrm{ns}$ 
is stretched to about $\tau_\text{rs:5-95\,\%}=150\,\mathrm{ns}$.

\begin{ack}
We are grateful to the inner tracker group of the HERA-B collaboration 
for providing several GEM foils. We thank C. Fiorini for providing the
preamplifier design.
\end{ack}

\end{document}